# TOWARDS A SERIOUS GAMES EVACUATION SIMULATOR

João Ribeiro[1], João Emílio Almeida[1,†], Rosaldo J. F. Rossetti[1,†], António Coelho[1,‡], António Leça Coelho[2]

[1]Department of Informatics Engineering
[†]LIACC – Laboratory of Artificial Intelligence and Computer Science
[‡]INESC TEC – INESC Technology and Science
Faculty of Engineering, University of Porto
Rua Roberto Frias, S/N, 4200-465, Porto, Portugal
{joao.pedro.ribeiro, joao.emilio.almeida, rossetti, acoelho}@fe.up.pt

[2]LNEC – National Laboratory of Civil Engineering
Av. Brasil, 101, 1700-066, Lisboa, Portugal
alcoelho@lnec.pt

**ABSTRACT**

The evacuation of complex buildings is a challenge under any circumstances. Fire drills are a way of training and validating evacuation plans. However, sometimes these plans are not taken seriously by their participants. It is also difficult to have the financial and time resources required. In this scenario, serious games can be used as a tool for training, planning and evaluating emergency plans. In this paper a prototype of a serious games evacuation simulator is presented. To make the environment as realistic as possible, 3D models were made using Blender and loaded onto Unity3D, a popular game engine. This framework provided us with the appropriate simulation environment. Some experiences were made and results show that this tool has potential for practitioners and planners to use it for training building occupants.

**INTRODUCTION**

The problem of evacuation from large facilities during an emergency or disaster has been addressed by researchers and practitioners in recent years. Real-world fire drills lack the realistic atmosphere of the emergency situation. Typically, the scenario is set up with the help of fire consultants and experts in the field, and the evacuation procedures follow some predefined rules and participants are expected to proceed accordingly.

In this paper, Serious Games (SG) are proposed as a means to overcome such drawbacks, since immersion into the emergency scenario artificially created using computer videogames is easier to accomplish. Also, the commitment of players, due to the excitement of using computer digital games, is expected to achieve better results than the traditional approaches.

In this paper the concept of serious games is used to build an evacuation simulator as an attempt to address some of the issues that were identified in real-world fire drills. It is our intention to improve the way people participate in such experiments enhancing their experience in many different ways. We have adapted and customised the environment of a game engine, in this case Unity3D, to support simulation features that enabled users to be tracked and assessed while playing. To test our approach and demonstrate its feasibility, we have carried out preliminary experiences with our prototype, in which subjects using the game environment were asked to evacuate a building in the case of fire.

The remaining part of this paper is organised as follows. We start by briefly presenting some related concepts that concern this project, such as pedestrian simulation and serious games. We then discuss on applying serious games to evacuation training, following the presentation and formalisation of our problem. We propose the approach implemented in this paper and suggest a preliminary experiment using our prototype. Some results are also discussed, after which we finally draw some conclusions and give clues of some further steps in this research.

**BACKGROUND AND RELATED WORK**

**Pedestrian simulators**

There are three main reasons for developing pedestrian computer simulations: i) to test scientific theories and hypotheses; ii) to assess design strategies; iii) to recreate the phenomena about which we want to theorize (Pan et al., 2007). Pedestrian flow management demands the correct representation of both the collective as well as the individual (Hoogendoorn et al., 2004). Timmermans et al. (2009) argue that understanding the pedestrian decision-making and movement is of critical importance to develop valid pedestrian models.

According to (Teknomo, 2002), pedestrian studies can be divided in two phases, namely data collection and data analysis. Whereas the former focus on characteristics such as speed, movement and path-planning, the latter is instead related to understanding how pedestrians behave. Predicting the movement of crowds (macroscopic level) or individual pedestrian actions (microscopic level) is the main goal of pedestrian simulation. For the macroscopic level, hydraulic or gas models are used (Santos and Aguirre, 2004). Microscopic models are based on behavioural approaches, in which entities are described individually (Castle et al., 2007). Traditional models, however, are mainly tested and validated through direct observations, techniques based on photography, as well as time-lapse films (Coelho, 1997; Helbing et al., 2001; Qingge et al., 2007) and also by stated preferences questionnaires (Cordeiro et al, 2011).

In such models it is possible to verify certain phenomena such as herding or flocking that happen due to people following other individuals instinctively. However, in conditions of low visibility or little knowledge of the surroundings this can provoke flocks of wandering people, contributing to the panic and confusion of the whole group, which is also a social reaction rather to be avoided if possible (Reynolds, 1987). Kuligowski proposes a model to mimic the human behavioural process during evacuation from buildings. Social science studies are needed to develop these theories, which could then yield more realistic results leading to safer and more efficient building design (Kuligowski, 2008, 2011).

Although many approaches exist to virtually simulate the behaviour of crowds with varying levels of realism, three models seem to be the most used (Heïgeas et al., 2003; Santos and Aguirre, 2004; Pelechano et al., 2007; Pretto, 2011). Cellular Automata Models (Neumann, 1966, Beyer et al., 1985) treat individuals as separate objects in an area divided into the so-called cells. Forces-based Models use mathematical formulaes to calculate the position variations of individual elements through the application of forces (its most explored subtypes consider Magnetic Forces and Social Forces). Finally, in Artificial Intelligence (AI) based Models, the decisions are made by individuals that compose the crowd on an autonomous basis. This sort of structure very much resembles a society of several interacting entities and has inspired much research in the Social Sciences (Kuligowski et al., 2010; Almeida et al., 2011).

**The Serious Games Concept**

Serious Games has gained a great prominence in the Digital Games field within the last decade, using appealing software with high-definition graphics and state-of-the-art gaming technology. It presents a great potential of application in a wide range of domains, naturally including social simulation.

Contrary to the primary purpose of entertainment in traditional digital games, SG are designed for the purpose of solving a problem. Although they are indeed expected to be entertaining, their main purpose is rather serious with respect to the outcomes reflected in changes to the player behavior (Frey et al., 2007; McGonigal, 2011).

According to (Hays, 2005), a game is an artificially constructed, competitive activity with a specific goal, a set of rules and constraints that is located in a specific context. Serious Games refer to video games whose application is focused on supporting activities such as education, training, health, advertising, or social change. A few benefits from combining them with other training activities include (Freitas, 2006): the learners' motivation is higher; completion rates are higher; possibility of accepting new learners; possibility of creating collaborative activities; learn through doing and acquiring experience. Other aspects that draw video game players' attention are fantasy elements, challenging situations and the ability to keep them curious about the outcomes of their possible actions (Kirriemuir et al., 2004). Serious Games can be classified in five categories: Edutainment, Advergaming, Edumarket Games, Political Games and Training and Simulation Games (Alvarez et al., 2007).

Bearing in mind the aforementioned characteristics of SG-based frameworks, we expect to contribute to the creation of the next-generation pedestrian simulators.

**A SERIOUS GAMES EVACUATION SIMULATOR**

The Serious Games Evacuation Simulator proposed in this research is based on the Unity3D game engine, that was selected due to its characteristics, among them: i) powerful graphical interface that allows visual object placement and property changing during runtime (especially useful to rapidly create new scenarios from existing models and assets and quick tweaking of script variables); ii) the ability to develop code in JavaScript, C# or Boo; iii) simple project deployment for multiple platforms without additional configuration, including for instance the Web (which makes it possible to run the game on a Web browser). Detailed characteristics of the implemented environment are presented below.

**Combining Simulation and Serious Games**

By starting the application the user gains control over the player character. Its aim will be to evacuate the building in the shortest time possible. The User Interface displays the elapsed time, which starts counting as soon as the player presses the "start fire simulation" key (illustrated in Figure 1).

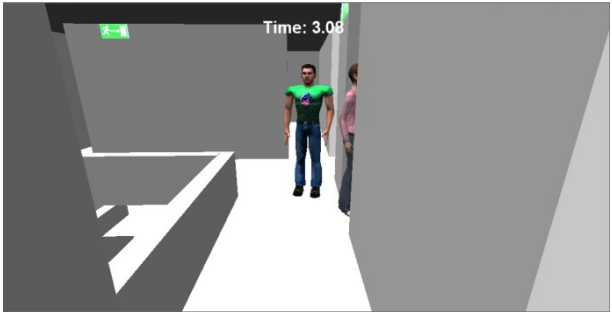

Figures 1: Gameplay example

**The game genre – First Person Shooter**

First Person Shooters (FPS) are characterised by placing players in a 3D virtual world which is seen through the eyes of an avatar. This attempts to recreate the experience of the user being physically there and exploring their surroundings.

The controls for this game follow the common standards for the FPS genre, using a combination of keyboard and mouse to move the player around the environment. The complete action mapping is as follows:

- **Mouse movement** - camera control, i.e. where the player is looking at;
- **W** - move forward;
- **S** - move backwards;
- **A** - move to the left;
- **D** - move to the right;
- **Space bar** - jump;
- **O** - start fire simulation.

**Game scenarios**

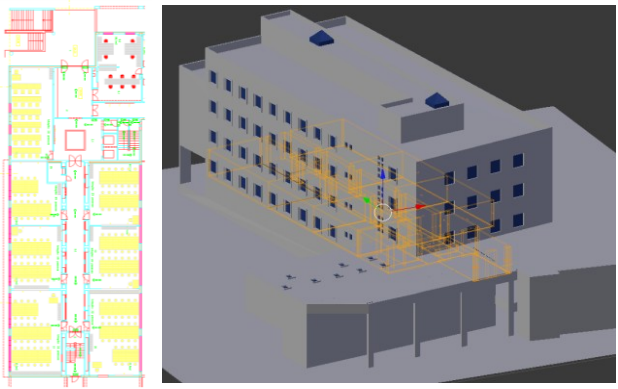

Figure 2 DEI plan and 3D representation

The environment is prepared to support various scenarios modelled in 3D. For the trial described in this paper, a single simulation scenario was considered. It takes place in FEUP's Informatics Engineering Department (DEI). A model of the FEUP campus was used, focusing only on one of the buildings where our research laboratory is located. As a virtual representation of the outside already existed, it was only necessary to create its interiors. This task was handled in Blender and used the official plans in order to recreate it as real as possible in terms of topology, dimensions, scale and proportions. Images of the plans and the 3D model are presented in Figure 2.

The player starts in a predefined room and, upon starting the evacuation event, a fire appears in a random room and the alarm sounds. At this very moment the timer starts. The player must then traverse the building in order to go to the outside as quickly as possible, choosing from one of the two possible exits. Several emergency signs are in place in order to help the player identify the nearest exit.

**Challenges, Rules and Scoring Systems**

The main challenge involved in the evacuation of a building comes from identifying the exact location of the nearest exit and how to get there. Also to consider is that computer-controlled agents are present and trying to evacuate the building at the same time, possibly clogging the passage and delaying the player.
After starting, fire keeps spreading to adjacent areas in small intervals of time; as fire is not surmountable, this can eventually constitute another obstacle and forces the player to look for a different exit route.
At the current stage, the score given to a player is solely based on the time taken to evacuate the building – meaning that the lower the score, the better. Whether the player picked the nearest exit or not is inherently reflected in the time taken to reach the outside.

**Model calibration**

Calibration is an important issue to assure the validity of the model. For this purpose, three different paths were considered, named P1, P2 and P3. One in a straight line (P1), two involving taking sets of stairs. Of these latter two, one involved taking the nearest exit from the building (P2) and another the farthest one (P3). These paths were measured using the AutoCAD plans for the building.

The comparison was made between data collected from real evacuations and from the game. The real times were measured with a stopwatch while traversing the paths, whereas for the game times the clock in the interface was used. It is also worth noticing that the adult profile of (1.5 m/s) was used in the game. Subject's speed values were calculated from the measured distance and time taken. Error values were calculated according to the equation:

$$1 - \frac{GameTime}{RealTime} \qquad (1)$$

The values for distances and times are registered in Table 1.

Table 1: Model Validation

|                     | P1    | P2    | P3    |
|---------------------|-------|-------|-------|
| Distance (m)        | 24    | 31    | 72    |
| Real Time (s)       | 17.53 | 21.50 | 55.91 |
| Subject's Speed (m/s)| 1.34 | 1.44  | 1.29  |
| Game Time (s)       | 15.86 | 19.28 | 48.08 |
| Error (%)           | 9.53  | 10.33 | 14.00 |

One aspect to notice is that subject's speed is consistent at around 1.3 and 1.4 m/s. Thus, subject's times would always be longer than the ones registered in the game, as the player moves at 1.5 m/s. It is also worth considering that the error is higher for routes involving stairs. This was also expected as the player's speed does not decrease when taking stairs, which in turn is verified in reality.

## EXPERIMENTAL SETUP

### Description

All subjects were divided according to their previous knowledge of the building. Each had to try to reach a safe exit to the outside as quickly as possible. Besides the time, it was expected that players would select the nearest emergency exit, just outside of the laboratory set as starting point, instead of using the normal way which is longer. Users were tested individually so not to spoil the experience to each other regarding details of their chosen routes. Tests were also performed only once in order to capture first reactions to the game experience and its controls.

### Population Sample

A total of 30 subjects were selected as sample to test the developed prototype. These testers can be classified according to the following parameters:

- Regular video game player - Yes or No;
- Familiar with the building -Yes or No.

An attempt was made to equalise these variables, as well as age and gender, so as to receive as many different experiences as possible and maintain a balance among categories. The distribution is shown in Table 2.

Table 2: User Times – Results by Categories

|                          |     | Regular Video Game Player | |
|--------------------------|-----|-----|-----|
|                          |     | Yes | No  |
| Previous Knowledge of the Building | Yes | 8   | 6   |
|                          | No  | 5   | 11  |

### Test Setup

Each subject could play only once. Some time was given to the user so as to get acquainted with the keyboard and mouse controls. Players with no previous knowledge of the building were taken to the lab where they had to escape from. The purpose was to show, like a regular visitor (for instance a student or foreign professor) the normal way, from the building entrance, up to the first floor and end of the corridor, where the laboratory is located. After the siren signed, the player was instructed to leave the building following the emergency signs leading to the nearest exit.

### Preliminary results

Intuitively it was expected that all subjects would selected the nearest exit available. However, some of the players misbehaved according to these expectations and chose the longer way out. These testers can be classified according to the following parameters, whose distribution is shown in Table 3.

Table 3: Experiment Results

| Was the nearest exit chosen?         | Y  | N  |
|--------------------------------------|----|----|
| Previous knowledge of the building   | 11 | 2  |
| No previous knowledge of the building| 6  | 11 |

From the analysis of Table 3, it is possible to conclude that users with previous knowledge of the building were aware of the emergency exit and used it. Nevertheless, 2 of them (aprox. 15% - 2 out of 13) missed it and used the longer way out. The remaining players, only 6 out of 17 (aprox. 35%) chose to exit using the emergency way, whilst the remaining 11 of that group followed the same way they were shown initially to get to the starting point of this experience.

## CONCLUSION AND FUTURE WORK

This work explores the concept of serious games as an important asset to aid and improve traditional fire drills. The contribution of this work can then be considered two-fold. First we extended a popular game engine to implement a pedestrian simulator to study evacuation dynamics. Second, our approach provided an appropriate environment to test with and influence behaviour of egresses of a building in hazardous situations, such as fires.

It also addresses the common notion that people tend to leave buildings using the same way they use to get into it, unless they are told otherwise. This was highlighted by the experiment in which approximately 65% of players without previous knowledge of the building missed the emergency exit and signage, following the longer but more intuitive path to exit the building.

It is important to bear in mind that this framework does neither completely replace nor avoid the need for in-site drills to train people for emergency situations, such as with the prospect of fire in an office building or school. Nonetheless, game environments can be very attractive in many different ways, and have proven to be an invaluable tool for training. Additionally, this approach is built upon the potential of such a concept to ease and improve the understanding of human behaviour in such situations, as subjects are monitored during their playing the game and some performance measures are logged to be further analysed later on.

We have implemented our prototype on the basis of a popular game engine, namely Unity3D, which provided us with a customisable framework and allowed us to feature the game virtual environment with characteristics of a serious game platform. We invited some subjects to use the game and collected some preliminary results that demonstrated the viability of the approach. We have then conceived a methodology which is both instrumental as an aid to train people and an invaluable instrument to help practitioners and scientists to better understand group behaviour and the social phenomenon in a vast range of circumstances.

The very next steps in this research include the improvement of the prototype featuring it with tools for rapidly setting up simulation environments from CAD blueprints of buildings. We also intend to include other performance measures to study individual and social behaviour in circumstances other the hazardous scenarios. Ultimately, this framework is also expected to be used as an imperative decision support tool, providing necessary and additional insights into evacuation plans, building layouts, and other design criteria to enhance places where people usually gather and interact rather socially, such as shopping malls, stadiums, airports, and so on.

## ACKNOWLEDGMENT

This project has been partially supported by FCT (Fundação para a Ciência e a Tecnologia), the Portuguese Agency for R&D, under grant SFRH/BD/72946/2010.

**AUTHOR BIOGRAPHIES**

**JOAO PEDRO RIBEIRO** concluded his MSc in Informatics and Computing Engineering in 2012, from Faculty of Engineering, University of Porto, Portugal. He specialised in Digital Games development and Artificial Intelligence, combining the concepts of multi-agent systems and serious games. He can be reached by e-mail at: `joao.pedro.ribeiro@fe.up.pt`.

**JOAO EMILIO ALMEIDA** holds a BSc in Informatics (1988), and MSc in Fire Safety Engineering (2008). He is currently reading for a PhD in Informatics Engineering at the Faculty of Engineering, University of Porto, Portugal, and a researcher at LIACC. He has co-authored many fire safety projects for complex buildings such as schools, hospitals and commercial centres. His areas of interest include Serious Games, Artificial Intelligence, and multi-agent systems; more specifically he is interested in validation methodologies for pedestrian and social simulation models. His e-mail is `joao.emilio.almeida@fe.up.pt`.

**ROSALDO ROSSETTI** is an Assistant Professor with the Department of Informatics Engineering at the University of Porto, Portugal. He is also a Research Fellow in the Laboratory of Artificial Intelligence and Computer Science (LIACC) at the same University. Dr. Rossetti is a member of the Board of Governors of IEEE Intelligent Transportation Systems Society (IEEE ITSS) and a co-chair of the Technical Activities sub-committee on Artificial Transportation Systems and Simulation of IEEE ITSS. His areas of interest include Artificial Intelligence and agent-based modelling and simulation for the analysis and engineering of complex systems and optimisation. His e-mail is `rossetti@fe.up.pt` and his Web page can be found at `http://www.fe.up.pt.com/~rossetti/`.

**ANTONIO COELHO** was born in 1971, in Porto, Portugal, and is currently an Assistant Professor at the Informatics Engineering Department of the Faculty of Engineering, University of Porto, where he teaches in the areas of Computer Graphics, Programming and Digital Games. He is also a Research Fellow at INESC TEC (INESC Technology and Science). His e-mail is `acoelho@fe.up.pt`.

**A. LEÇA COELHO** holds both the Electrotechnical and Civil Engineering degrees, as well as a Master's and PhD in Civil Engineering. He is currently a Principal Researcher with Habilitation at LNEC. His areas of interest include fire safety and risk analysis. He can be reached by e-mail at `alcoelho@lnec.pt`.